\documentstyle[11pt]{article}
\def\emptyc{$\left.\right.$}

\def\FigCaption#1#2{
 \vspace{2mm}
 \vbox{
   \noindent{\large\bf Fig.\ #1}
   \begin{quotation}
   \noindent #2
   \end{quotation}
  }
}

\font\msamstex=msam10% scaled \magstep 1
\def\grtsim {\,\mbox{{\msamstex \char 38}}\,}

\begin{document}

\title{Random copolymer:\\Gaussian variational approach.}
\author{A.~Moskalenko,~Yu.A.~Kuznetsov,~ K.A.~Dawson}
\date{{\small\it Theory and Computation Group,\\ 
Centre for Colloid Science and Biomaterials,\\
Department of Chemistry, University College Dublin,\\
Belfield, Dublin 4, Ireland\\
%e-mail: andrei@fiachra.ucd.ie
}
\emptyc \\
{\normalsize\rm \today}}
\maketitle
%\footnotetext[1]{e-mail: andrei@fiachra.ucd.ie}
\begin{abstract}
We study the phase transitions of a random copolymer chain 
with quenched disorder. We apply a replica variational 
approach based on a Gaussian trial Hamiltonian in terms of 
the correlation functions of monomer Fourier coordinates.
This allows us to study collapse, phase separation and 
freezing transitions within the same mean field theory.
The effective free energy of the system is derived 
analytically and analysed numerically. Such quantities as 
the radius of gyration or the average value of the overlap 
between different replicas are treated as observables and 
evaluated by introducing appropriate external fields to 
the Hamiltonian. We obtain the phase diagram and show that 
this system exhibits a scale dependent freezing transition. 
The correlations between replicas appear at different 
length scales as the temperature decreases. This 
indicates the existence of the topological frustration.
\end{abstract}
\section{Introduction}
Various models of random copolymers have been extensively studied using 
theoretical
\cite{Obukhov-86,Garel-88,Shakhnovich-89,Fredrickson-92,Vilgis-94,Garel-95} 
and  numerical 
\cite{Chan-91,Grosberg,Socci-95} methods.
It has been found that even simple models exhibit a rich variety of 
conformational states 
leading to quite a complicated phase diagram. 

Most of the theoretical results have been 
obtained within the replica approach
\cite{Mezard-book} for directed polymers in random media 
\cite{Mezard-91,Edwards-88,Orland-95} 
and for self--interacting copolymers under the
assumption of the 
ground state dominance and constant density approximations
for Gaussian disorder \cite{Garel-88,Shakhnovich-89,Sfatos-93}.
Computer simulations \cite{Chan-91,Grosberg,Socci-95} 
have qualitatively confirmed 
some of the theoretical predictions.

In this paper we study the complete phase diagram of a random copolymer chain
with attractive and repulsive units. It is believed that
such studies have application in the field of protein folding and related 
subjects \cite{Bryngelson}.
Thus we apply a Gaussian variational approach based on a generic 
quadratic trial 
Hamiltonian in terms of monomer Fourier coordinates of replicas, 
which is an alternative to the standard one
using the density overlap function
as the replica order parameter and constant density approximation.  
A similar approach was applied 
to directed polymers and manifolds in random media \cite{Edwards-88,Mezard-91}.
The method has the advantage that it allows us to incorporate fluctuations 
of the density, determined
self--consistently, and to study collapse, phase separation
and glass transitions within the same mean field theory.
%We have calculated the effective free energy of the system and a number of 
%observables in the one--step Parisi scheme. 

Note that the self--consistent method has been applied 
to various other problems at
equilibrium \cite{Cloizeaux,Allegra} and in kinetics \cite{Timoshenko} of
homo-- and block copolymers and to directed polymers in random media
\cite{Edwards-88,Orland-95}. 
%\section{Method}

It is convenient to represent the Hamiltonian
as the sum of two parts, $H =H_{hom} + H_{dis}$.
The first is purely homopolymeric and 
the second is related to the quenched disorder. Thus, 
%%%%%%%%%%%%%%%%%%%%%%%%%%%%%%%%%%%%%
\begin{eqnarray}
\label{ham}
&&\beta H_{hom} = \frac{\kappa}{2}\int_0^N ds 
\left(\frac{d{\bf r}(s)}{ds}\right)^2 
\\
&&+ \sum_{m=1}^{\infty}\frac{u_{m+1}}{(m+1)!}\int_0^N \{ds_0...ds_m \} 
\prod_{i=1}^{m}
\delta ({\bf r}({s_0})-{\bf r}({s_i})) \,,
\nonumber
\end{eqnarray}
%%%%%%%%%%%%%%%%%%%%%%%%%%%%%%%%%%%%
\begin{eqnarray}
\label{hdis}
H_{dis}=\frac{1}{2}\int_0^N\int_0^N dsds'\, u(\lambda_{s},\lambda_{s'})\, 
\delta ({\bf r}({s})-{\bf r}({s'})) \,,
\end{eqnarray}
where the connectivity constant is $\kappa =3l^{-2}$, 
with $l$ being the 
bond length. Here the $u_{n}$ denote the virial coefficients
and the curly braces indicate that the terms
with coinciding indices are excluded.
The matrix of two--body interactions  may be written 
\cite{Obukhov-86}
via the quenched random variables $\{\lambda_{s}\}$,  
\begin{eqnarray}
\label{matrix}
u_2(\lambda_s,\lambda_{s'})=v+a(\lambda_s+\lambda_{s'})
+\chi\lambda_{s}\lambda_{s'},
\end{eqnarray}
where the parameters $v$, $a$ and $\chi$ may be expressed in terms of,
for example, 
$u_{AA},\ u_{BB}$ and $u_{AB}$ in the case of only two monomer types. 
The average over the quenched disorder is performed by introducing
replicas \cite{Mezard-book}
\begin{eqnarray}
%\label{}
\beta F_{quench}&=&
-\overline{\log Z(\{\lambda_s\})}
=-\lim_{n \to 0}
n^{-1}\cdot \biggl(\overline{ Z^n(\{\lambda_s\})} 
- 1\biggr)\,,
\\
&&\overline{ Z^n(\{\lambda_s\})}=\int D{\bf r}_i^a
\exp{(-\beta H_n)}
\end{eqnarray}
where $Z$, $H_n$ are respectively the partition function of the system and
the effective Hamiltonian in the replica space. Here
the bar stands for the average over disorder.

For simplicity we assume the distribution of $\{\lambda_s\}$ to be
Gaussian with mean value $\lambda_0$ and variance $\Delta^2$,
\begin{eqnarray}
\label{}
P(\{\lambda_s\})=\prod_s\frac{1}{(2\pi\Delta^2)^{1/2}}
\exp{\left( -\frac{(\lambda_s-\lambda_0)^2}{2\Delta^2}\right)}.
\end{eqnarray}
However other distributions, for example the binary distribution, can
be considered. 
Averaging over the quenched disorder
produces a compact expression for the effective Hamiltonian in the 
replica space,
\begin{eqnarray}
\label{heffect}
\beta H_n=\beta H^n_{hom}
- \frac{{(\beta\Delta a)}^2}{2!}
\int_0^N dsds_1ds_2\sum_{ab}
\delta ({\bf r}^a(s,s_1))
\delta ({\bf r}^b(s,s_2))\,,
\end{eqnarray}
\begin{eqnarray}
{\bf r}^a(s,s_1)\equiv {\bf r}^a(s)-{\bf r}^a(s_1)\,, 
\end{eqnarray}
where we assume $\chi =0$, $\lambda_0 =0$
and that the Hamiltonian is regularized by introduction of
a short distance cutoff.
This choice of parameters in the matrix of interactions corresponds 
to the amphiphilic
model, in which monomers of $A$--type attract each other and
monomers of $B$--type repel each other. We note here that 
the case $\chi\not =0$ requires
some additional treatment, 
for example application of the perturbation expansion
of the matrices $M^{-1}$ and $\mbox{log}\mbox{$M$}$,
where 
\begin{eqnarray}
%\label{}
\mbox{$M(s,s')$}=\delta (s-s') + 
\beta\Delta^2\chi \sum_a\delta ({\bf r}^a(s)-{\bf r}^a(s'))\,.
\end{eqnarray}

\section{Variational approach}

It is useful to introduce the Fourier coordinates of the monomer positions.
For ring polymers the appropriate Fourier transform is defined as
\begin{eqnarray}
\label{fourier}
{\bf r}_{q}=\frac{1}{N}\int_0^N d{s}\,
\exp{\biggl(\frac{-i2\pi sq}{N}\biggr)}{\bf r}(s)\,,\quad
{\bf r}(s)={\bf r}_{0}+\sum_{q=1}^{\infty}
\exp{\biggl(\frac{i2\pi sq}{N}\biggr)}{\bf r}_{q}\,,
\end{eqnarray}
and the generalization to the case of open chain is straightforward.
The standard Gibbs--Bogoliubov variational principle gives
the estimate for the free energy,
\begin{eqnarray}
\label{var_ast}
F_n=F_0+\langle (H_n-H_0) \rangle_0.
\end{eqnarray}
We choose the trial Hamiltonian as a quadratic form, which is
standard in the
polymer literature \cite{Mezard-91,Edwards-88}, but
with $q$--dependence of all variational parameters
\begin{eqnarray}
\label{trial}
\beta H_0=\frac{1}{2}\sum_{a,b}\sum_q V_{ab}(q){\bf r}_{-q}^a{\bf r}_q^b,
\end{eqnarray}
where the effective potential matrix is 
to be taken non-diagonal in the replica indices and 
diagonal in the chain index
due to the assumption of translational invariance along 
the chain after integrating out
the disorder.
We note here that formally one has to add
a small mass term $\mu\delta_{ab}$ to
the effective potential $V^{(q)}_{ab}$ in (\ref{trial}) for regularisation
\cite{Mezard-91}. The limit $\mu \to 0$ is assumed to be taken at the
final stage of calculations.
 
Averaging (\ref{heffect}) over the statistical ensemble 
(\ref{trial}) we obtain the effective 
free energy of the following
form
\begin{eqnarray}
\label{effective free energy Appendix}
\beta F_n&=&-\frac{3}{2}\sum_{q}\, {\mbox{Tr log}} {\hat{{\cal F}}_q}
+\frac{6\kappa\pi^2}{N}\sum_a\sum_{q}\, q^2{\cal F}^{aa}_q
\nonumber\\
&+&\sum_{m=1}^{\infty}
\frac{u_{m+1}-3 \delta_{m,2}\tilde{\Delta}^2}{(m+1)!(2\pi)^{3m/2}}
\sum_a\int { \{ ds_0...ds_m \} }
{(X^{a...a}_{1...m}) }^{-3/2}
\\
&-&\frac{\tilde{\Delta}^2}{2(2\pi)^3}\int ds_1ds_2ds_3\sum_{a\not =b}
(D^{aa}_{s_1s_2}D^{bb}_{s_2s_3}-(D^{ab}_{s_1s_2s_3})^2 )^{-3/2}\,.
\nonumber
\end{eqnarray}
Here we introduced $\tilde{\Delta}=\beta{\Delta}a$ and matrices of spatial 
$D^{ab}_{s_1s_2}=\frac{1}{3}\langle ({\bf r}^a(s_1)-{\bf r}^b(s_2))^2\rangle_0$
%$\hat{D}_{ss'}$ 
and Fourier 
${{\cal F}^{ab}_q}=\frac{1}{3}
\langle {\bf r}_{-q}^a{\bf r}_q^b \rangle_0$
monomer correlations
\begin{eqnarray}
\label{D}
X_{1...n}=\mbox{det}_{n\times n}{D}_{l:k}\,,
{D}_{l:k}={D}^{aa}_{s_ls_0s_k}\,,
2D^{ab}_{s_1s_2s_3}=D^{ab}_{s_1s_2}+D^{ab}_{s_2s_3}-D^{ab}_{s_1s_3}\,.
\nonumber
\end{eqnarray}

The procedure of performing the limit
$n\to 0$    
%%%%%%%%%%%%%%%%%%%%%%%%%%%%% RSB PARISI SCHEME %%%%%%%%%%%%%%%%%%%%
is well known from studies of spin glass systems \cite{Mezard-book,Mezard-91}. 
To do so we assume that ${{\cal F}^{ab}_q}$ is a Parisi--type hierarchical 
matrix parametrised by
$\{\tilde{{\cal F}}_q,{\cal F}_q(u) \}$, $u\in[0,1]$, where 
$\tilde{{\cal F}}_q$ and ${\cal F}_q(u)$ are correlation functions for
monomers from the same replica and from different replicas respectively.
%%%%%%%%%%%%%%%%%%%%%%%%%%%%%%%%%%%%%%%%%%%%%%%%%%%%%%%%%%%%%%%% 
Finally,
we obtain %from Eq. (\ref{var_ast}) 
the effective free
energy per replica in the limit $n\to 0$, which is
convenient to represent as a sum of two terms,
\begin{eqnarray}
\label{effen}
\beta F_{eff}=
\lim_{n\to 0}(\beta F_{n}/n) =\beta N \biggl(f_0[\tilde{{\cal F}}_q] 
+ f_1[\tilde{{\cal F}}_q, {\cal F}_q(u)]\biggr)\,.
\end{eqnarray}
The first term, $f_0$, coincides with
the annealed effective free energy, 
%except for the term ${\cal Q}^{(0)}_{12}$,
%which is defined below. 
%In fact, $f_0$ is the only term in the free 
%energy (\ref{effen}), if one uses the diagonal trial Hamiltonian
%(\ref{trial}), 
\begin{eqnarray}
\label{fann}
&&\beta N f_0 =-\frac{3}{2}\sum_{q}\, {\mbox{Tr log}} {\tilde{{\cal F}}_q}
+\frac{6\kappa\pi^2}{N}\sum_{q}\, q^2{\tilde{\cal F}}_q
\nonumber\\
&&+\sum_{l=1}^{\infty}
\frac{u_{l+1}-3 \delta_{l,2}\tilde{\Delta}^2}{(l+1)!(2\pi)^{3l/2}}
\int { \{ ds_0..ds_l \} }
{X_{1..l} }^{-3/2}\,, 
\end{eqnarray}
where $X_{1...m}$ are determinants of $m\times m$-matrices
\begin{eqnarray}
\label{Pi}
X_{1...m}=\mbox{det}_{m\times m}\tilde{D}_{l:k}\,,\quad
\tilde{D}_{l:k}=\tilde{{\cal D}}_{i_li_0i_k}=\sum_q d^{(q)}_{i_li_0i_k}
\tilde{{\cal F}}_q\,.
\nonumber
\end{eqnarray} 
The second term, $f_1$, is determined purely by the quenched disorder,
%Using the standard technique we write
\begin{eqnarray}
\label{quencheff}
\beta Nf_1 &=& -\frac{3}{2}\sum_q\biggl[\mbox{Log}\ (1-\langle {\cal R}_q\rangle)
+\frac{{\cal R}_q(0)}{1-\langle{\cal R}_q\rangle}
\nonumber\\
&-&\int^1_0\frac{du}{u^2}\mbox{Log}\ 
\frac{1-\langle{\cal R}_q\rangle-[{\cal R}_q](u)}{1-\langle{\cal R}_q\rangle}\biggr]
\\
&+& \frac{{\tilde\Delta}^2}{2!(2\pi)^{6/2}}
\int d{s_0}d{s_1}d{s_2}\int^1_0 du \, [{\tilde{\cal D}_{s_0s_1}}
\tilde{\cal D}_{s_1s_2}-
{{\cal D}^2_{s_0s_1s_2}}(u)]^{-3/2}
\nonumber
\end{eqnarray}
where by definition for any arbitrary matrix $a$:
\begin{eqnarray}
&&\langle a\rangle = 
\int^1_0 du \, a(u) \,,~~[a](u)=-\int^u_0 dv \,a(v) +ua(u)\,,\\
\mbox{and} \nonumber\\ 
&&
0\leq {\cal R}_q(u) ={\cal F}_q(u) / \tilde{{\cal F}}_q <1 \,.
\end{eqnarray}
Here we introduce ${\cal R}_q(u)$, which 
gives the measure of correlations between different replicas relatively to 
correlations in the same replica and 
allows the division on ``annealed'' and ``quenched'' parts (\ref{effen}).    
%%%%%%%%%%%%%%%%%%%%%))))))))))))%%%%%%%%%%%%%%%%%%%%%%%%%%%%%%%
The effective free energy has to be minimized with respect to
$\tilde{{\cal F}}_q$ and maximized with respect to
${\cal F}_q(u)$
with the constraints $0\leq {\cal R}_q(u)<1$.
The latter physically means that monomer coordinates' correlations 
between different replicas
are always weaker than those in the same replica.
Alternatively these constraints can be derived if we require
the matrices $V_{ab}(q)$ and ${\cal F}^{ab}_q$ to have positive eigenvalues
for arbitrary integer $n$, $n\geq 1$, and all non--diagonal elements of
the matrices $V_{ab}(q)$ to be negative to match the sign
of the corresponding non--diagonal terms in the effective Hamiltonian.

Usually one is interested in such quantities as radius of 
gyration $R_g$, end--to--end distance, phase separation
order parameters or the average value $Q$ of
an overlap between different replicas $Q_{ab}$, for example,
%The micro--phase separation order parameter can be defined as 
\begin{eqnarray}
\label{orderpar}
&&R_g^2=1/2 N^{-2}\int d{s}d{s'}\langle
\overline{({\bf r}(s)-{\bf r}(s'))^2}\rangle_0\,,
\nonumber\\
&&Q_{ab}=l^3N^{-1}\int d{s}\langle
\overline{\delta({\bf r}^a(s) - {\bf r}^b(s) )}\rangle_0\,,
\\
&&\delta R^2_g= 1/2 N^{-2}\int d{s}d{s'}\langle
\overline{\lambda_s({\bf r}(s)-{\bf r}(s'))^2}\rangle_0\,.
\nonumber
\end{eqnarray}
All these can be treated as observables and evaluated
by introducing appropriate external fields to the partition
function. The model we consider
exhibits a strong tendency for phase--separation. Therefore, the
parameter $\Psi=\delta R^2_g/R^2_g$ can be considered as
a measure of phase separation, where $\delta R^2_g$ is the
difference of the square radii of gyration in  
the case of two letter copolymers with equal concentrations of both monomer types,
$\delta R^2_g=R^2_{gA}-R^2_{gB}$. Introducing the term $\mu\delta R^2_g$
into the Hamiltonian one obtains
\begin{eqnarray}
&&\delta R^2_g = -\lim_{\mu ,n\to 0}\frac{\partial}{\partial\mu}
\frac{\partial}{\partial n}\biggl\langle Z^n(\mu) \biggr\rangle
\\
&&=\frac{3}{4}\frac{{\tilde\Delta}\cdot\Delta}{(2\pi)^{3/2}N^2}
\int d{s_0}d{s_1}d{s_2}
\frac{{\tilde{\cal D}_{s_0s_1s_2}}^2
-\langle {\cal D}_{s_0s_1s_2}^{~2}\rangle}
{\tilde{\cal D}_{s_1s_2}^{5/2}}
\nonumber
\end{eqnarray}

To determine the freezing transition one may study the change of 
the average overlap among replicas $Q$ as well as the eigenvalues
of the matrix of the second derivatives of the free energy (\ref{effen})
with respect to all ${\tilde{\cal F}}_q$ and ${\cal F}_q(u)$, 
Hessian $\hat{T}$.
The solution we are looking for is a saddle point with the positive eigenvalues
of $\hat{T}$ corresponding to ${\tilde{\cal F}}_q$ and negative ones
corresponding to ${\cal F}_q(u)$. As one can see from the numerical analysis,
the solution with all ${\cal F}_q(u)=0$, no correlations among replicas, 
is not a saddle point of the type mentioned above for 
the variance of the disorder ${\tilde\Delta}$ larger than some particular
critical value ${\tilde\Delta}_f$. The latter can vary with a change
in the values of the virial coefficients. 
%%%%%%%%%%%%%%%%%%%%%% ONE STEP %%%%%%%%%%%%%%%%%%%%%%%%%%%%%%%%%%%%%%%%%%
%%%%%%%%%%%%%%%%%%%%%%%%%%%%%%%%%%%%%%%%%%%%%%%%%%%%%%%%%%%%%%%%%%%%%%%%%%%%

\section{Numerical Analysis}

%\font\msamstex=msam10% scaled \magstep 1
%\def\grtsim {\,\mbox{{\msamstex \char 38}}\,}
To analyze system numerically one has to substitute
integration by summation
in (\ref{effen}), see Appendix A for details.
In principle 
we can evaluate the effective free energy for any finite number of steps in
the Parisi scheme.
In practice this means the introduction of a 
large number of variational parameters. If one has a polymer of length 
$N$ monomers and applies the $k$--step Parisi scheme, then one has to
analyze a set of
approximately $N(k+1)$ nonlinear algebraic equations. 
Therefore, we apply a minimal
one--step Parisi scheme, which still incorporates all the main features of 
the exact solution and
allows one to locate the onset of the transition to phases with 
broken replica symmetry if such exists. Thus we approximate 
${\cal F}_q(u)$ by the step function 
\begin{eqnarray}
{\cal F}_q(u)=  \left\{\begin{array}{clcr}
                 {\cal F}^{(0)}_q  & \mbox{$u<x$, correlations between replicas from different groups,} \\
                 {\cal F}^{(1)}_q  & \mbox{$u>x$, correlations between replicas from the same group,}
               \end{array}
           \right.          
\end{eqnarray}

in one--step approximation:

\begin{eqnarray}
\label{fquench}
&&\beta N f_1 
=-\frac{3}{2}\sum_q \left[\frac{1}{x} 
\log{\left(1 -(1-x){\cal R}^{(1)}_q-x{\cal R}^{(0)}_q\right)}\right]
\nonumber\\
&& -\frac{3}{2}\sum_q \left[
\frac{x-1}{x}\log\biggl( 1 -{\cal R}^{(1)}_q\biggr)+
\frac{{\cal R}^{(0)}_q}{1-(1-x){\cal R}^{(1)}_q-x{\cal R}^{(0)}_q}
\right]
%\\
\\
&&+
\frac{{\tilde\Delta}^2}{2!(2\pi)^{6/2}}
\sum_{ \{ ijk \} }
\biggl[
(1-x){Q^{(1)}_{ijk}}^{-3/2}
+x{Q^{(0)}_{ijk}}^{-3/2} %- {{\cal Q}^{(0)}_{12}}^{-3/2}
\biggr]\,,
\nonumber\\
&&Q^{(s)}_{ijk}={\tilde{\cal D}_{ij}}\tilde{\cal D}_{jk}-{{\cal D}_{ijk}^{(s)}}^{2}\,,
~{\cal D}_{ijk}^{(s)}=\sum_q d^{(q)}_{ijk}{\cal F}^{(s)}_q\,,\quad s=0,1\,.
\end{eqnarray}

In the following we restrict ourselves to the choice of
parameters: $k_B T = 1$, $l = 1$, which provide the scaling units of
energy and length; and we set $u_m = 0$ for $m\geq 5$. 
Note that the zero--modes ${\tilde{\cal F}}_0$ and ${\cal R}^{(0,1)}_0$
do not contribute to the mean energy of the system
describing diffusion of the center of mass of the polymer chain.

The system exhibits two different types of behavior above and below 
the freezing transition.
Above the freezing temperature, 
the monomer coordinates correlation function for monomers 
from the same replica, ${\tilde{\cal F}}_q$, 
closely coincides with that for
annealed disorder, whilst  
the monomer coordinates correlation functions for monomer from different replicas,
${\cal R}^{(0)}_q$ and ${\cal R}^{(1)}_q$,
possess only trivial solutions, namely: ${\cal R}^{(0)}_q = {\cal R}^{(1)}_q = 0$.
The coil and globular states are
similar to the homopolymer coil and liquid--like globule states respectively.
The trial Hamiltonian
is diagonal in the replica space and
the free energy differs from the annealed one by the term
$$
{\beta N f_1=\frac{{\tilde\Delta}^2}{2!(2\pi)^{6/2}}
\sum_{ \{ m_0m_1m_2 \} } ({\tilde D}_{m_0m_1}{\tilde D}_{m_0m_2})^{-3/2}} \,,
$$
which remains regular in the whole range of the parameter space 
and becomes negligibly small 
in the region of small ${\tilde\Delta}$.

The situation changes around some critical value of the dispersion of
disorder $\tilde{\Delta}_f$, at which point the solution yields nontrivial values
for the correlation functions 
${\cal R}^{(0)}_q$ and ${\cal R}^{(1)}_q$ and the parameter $x$. 
Thus, the onset of the freezing transition can be seen clearly
from the behavior of both $x$ and the set of ${\cal R}^{(0,1)}_q$, where they
start to change rapidly. 
%Further, the replica correlation function
%${\cal R}^{(1)}_1$ continues to
%increase asymptotically reaching the value ${\cal R}^{(1)}_q\rightarrow 1$
%when $\tilde{\Delta}\rightarrow\infty$. Nevertheless, the quantities
%$x$ and ${\cal R}^{(0)}_q$ may deviate from monotonic dependence as one
%can see from Fig.~1.
The typical behavior of the correlation functions for monomers 
from different replicas can be seen in Figs~1 and~2. 

Fig.~1 presents the correlation function
for monomers from the same group of replicas, ${\cal R}^{(1)}_{q=1}$,
 and from different groups
of replicas, ${\cal R}^{(0)}_{q=1}$,
and the parameter $x$ versus the dispersion of disorder ${\tilde\Delta}$
at fixed values of the other parameters in the free energy (\ref{effen}).
Strictly speaking the solution below freezing is valid only in the vicinity of the
transition line, though we draw it for a wide range of the dispersion
of disorder, keeping in mind that more steps in the Parisi scheme are required
to describe the properties of the frozen phase. 
After the freezing transition the coordinates of monomers from
different replicas become correlated which results in 
a decrease of the entropy of the system.
The symmetry among replicas is spontaneously broken, the correlations between
the positions of the monomers from the same group of replicas are higher 
than those between
the positions of the monomers from different groups.

Fig.~2 presents the profile of the replica's correlation 
functions ${\cal R}^{(0)}_{q}$ and ${\cal R}^{(1)}_{q}$ in the chain index $q$.
There are no correlations for large $q$, $q<N/2$ for ring polymers, 
between different replicas in 
the vicinity of the freezing transition  
although these become non--zero
as $\tilde{\Delta}$ increases. 
To locate the onset of
the freezing transition only 
correlations on the first mode, ${\cal R}^{(0)}_{1}$ 
and ${\cal R}^{(1)}_{1}$, and the parameter $x$ are needed. 
This clearly indicates the scale dependence of the freezing
due to the topological frustration in the system
and is in agreement with the main result of \cite{Thirumalai}.

The resulting phase diagram of the random copolymer is
represented in Fig.~3. It shows that
the collapse can occur in two different ways.
For small values of the variance of the two body interaction, ${\tilde\Delta}$, 
the coil collapses to the liquid--like globule as
the effective two body interaction changes from repulsion to attraction
and ${\tilde u}_2$ becomes negative.
For large ${\tilde\Delta}$ the effective three body interaction,
$(u_3)_{eff}=u_3-3{\tilde\Delta}^2$, 
changes sign from positive to negative and 
collapse occurs due to the three body effects \cite{Garel-95},
even when ${\tilde u}_2$ remains positive, but small.   
There is also a transition from the liquid--like globule to the frozen
globule %\cite{Sfatos-93} 
(line (F) in Fig.~2).
For very large ${\tilde \Delta}$ there is an indication 
that the copolymer chain collapses directly to the frozen phase and that 
the phase diagram may exhibit a multi-critical point where all three phases,
coil, liquid--like and frozen globule coexist.
%However, we have not established this as yet, within the current approximation.

The freezing transition line on the Fig 3 scales approximately
as $\tilde{\Delta}_f\sim |{u}_2|^{-\gamma}$ for
large negative ${\tilde u}_2$, where $\gamma\simeq 0.66\pm 0.03$.
%Practically, this ia a crossover region between two different regimes.
%For small $\tilde{\Delta}$ $\gamma$ is 
The freezing transition line is nearly independent of the third virial
coefficient in the region where ${u}_3 \ll {u}_4$ and increases
linearly in ${u}_3$ when ${u}_3 \grtsim {u}_4$.
In a more dense globule, i.e. given a bigger negative ${u}_2$
or smaller $\tilde{u}_3$, chains undergo a transition from a liquid--like
to a frozen globule at smaller values of ${\tilde\Delta}$.
%Practically, the exponent $\gamma$ is a crossover exponent 
Both the collapse transition and the freezing transition lines
are essentially independent of the size of the chain for sufficiently
large degree of polymerization $N$.

In Fig.~4 we present the phase separation parameter $\delta R^2_g$ versus
the dispersion $\tilde{\Delta}$,
\begin{eqnarray}
\delta R^2_g =
\frac{3}{4}\frac{{\tilde\Delta}\cdot\Delta}{(2\pi)^{3/2}N^2}
\sum_{i\{jk\}}
\frac{{\tilde{\cal D}_{ijk}}^2
-x{{\cal D}_{ijk}^{(0)}}^2 -(1-x){{\cal D}_{ijk}^{(1)}}^2}
{\tilde{\cal D}_{jk}^{5/2}} \,.
\nonumber
\end{eqnarray}
This parameter increases linearly in ${\tilde\Delta}$
in the liquid--like globule phase, but remains constant in the frozen phase,
whilst $\Psi$ reaches a maximum at the freezing transition and starts 
to decrease slowly in the frozen phase due to the increase in the radius
of gyration $R_g$. This means that freezing prevents the system from further
phase separation and therefore, for example, the observable $\delta R^2_g$ 
and its first derivative with respect to ${\tilde\Delta}$ can also be used
to locate the onset of the freezing transition. 

%In Fig.~4 we exhibit the profile of replica correlation functions
%${\cal R}^{(0,1)}_q$ in the chain index $q$. Across the freezing
%transition the most long--range correlations ${\cal R}^{(0,1)}_1$ start to
%increase, whilst the short--range correlation functions remain zero even far
%below the freezing transition.

\section{Discussion and Conclusions}
In this paper we have applied a Gaussian variational approach
to study
the phase transitions of random copolymers.
The essential feature of this approach is that it
is based on a trial Hamiltonian in terms of monomer coordinates
rather than density variables.
For this reason it may be used to describe a polymer with arbitrary
Gaussian disorder across the whole range of the collapse transition.
Our equations reduce to the self--consistency 
equations for a homopolymer at equilibrium
as the variance of the second virial coefficient vanishes.
The most important result is that the freezing transition in the system is
scale dependent \cite{Thirumalai}.
 
We have performed a numerical 
analysis of the free energy in the one--step Parisi
scheme. We note here that, firstly, more steps in the Parisi scheme
may be required to describe the property of the frozen
phase and, secondly, the replica symmetrical solution
brings qualitatively the same results.

It is believed that the model we have considered
can provide a basis for understanding conformational states of
biomolecules such as proteins \cite{Bryngelson}. The results
for the freezing transition and the phase separation are in agreement
with those obtained for this model by other methods \cite{Kyoto}. 
We note that the results we have obtained are valid 
also for binary distribution of disorder, provided that 
higher virial coefficients
remain positive. However, the whole diagram in this case will
incorporate new features, e.g. the explicit dependence of
the freezing transition on the chemical composition.

We also note that one of the important drawbacks of the theory is that
the Gaussian method yields an incorrect Flory exponent for a chain
in a good solvent.
However, this problem can be resolved in different ways 
and we refer the reader to the discussion of this 
point in \cite{Kuznetsov-96}.

Finally we may comment that despite the limitations, 
the current approach offers one of the few consistent 
approaches to the global ``phase'' diagrams of
random amphiphilic copolymer chain \cite{Kyoto,Timosh-96}.

The authors thank A. Gorelov, E. Timoshenko, I. Moskalenko 
and D. Hegarty for comments. One of the authors (A.M.)
thanks the organizers of PeH 96, where part of this work 
was presented.  
%%%%%%%%%%%%%%%%% APPENDIX A %%%%%%%%%%%%%%%%%%%%%%%%%%%%%%%%%%%%%%%
\newpage
\section{Appendix A}
\setcounter{equation}{0}
\def\theequation{A.\arabic{equation}}

To perform a numerical analysis we have to substitute
integration by summation,
\begin{eqnarray}
\int_0^N d{\{s_1s_2...s_m\}}
\leftrightarrow
\sum_{\{i_1,i_2...i_m\}=0}^{N-1}\,, 
\frac{6\kappa\pi^2}{N}\sum_{q}\, q^2\tilde{{\cal F}}_q
\to6\kappa N\sum_{q=1}^{N-1} {\tilde{\cal F}}_q\sin^2 \frac{\pi q}{N}\,
\end{eqnarray}
where $\{i_1i_2...i_m\}$ indecates that indeces $i_1,i_2,...,i_m$ are
not allowed to coincide. 
Thus we can rewrite (\ref{fann}) and (\ref{quencheff}) 
\begin{eqnarray}
%\label{fann}
&&\beta N f_0 =-\frac{3}{2}\sum_q\log \tilde{{\cal F}}_q
+6\kappa N\sum_q {\tilde{\cal F}}_q\sin^2 \frac{\pi q}{N}
\nonumber\\
&&+ 
\sum_{n=1}^{\infty}
\frac{{u}_{n+1}-3\delta_{n+1,3}\tilde{\Delta}^2}{(n+1)!(2\pi)^{3n/2}}
\sum_{ \{ m_0...m_n \} }
{X_{1...n} }^{-3/2}\,, 
\end{eqnarray}
\begin{eqnarray}
%\label{quencheff}
\beta Nf_1 &=& -\frac{3}{2}\sum_q\biggl[\mbox{Log}\ (1-\langle {\cal R}_q\rangle)
+\frac{{\cal R}_q(0)}{1-\langle{\cal R}_q\rangle}
-\int^1_0\frac{du}{u^2}\mbox{Log}\ 
\frac{1-\langle{\cal R}_q\rangle-[{\cal R}_q](u)}{1-\langle{\cal R}_q\rangle}\biggr]
\nonumber\\
&+& \frac{{\tilde\Delta}^2}{2!(2\pi)^{6/2}}
\sum_{ \{ ijk \} }\int^1_0 du \, [{\tilde{\cal D}_{ij}}\tilde{\cal D}_{jk}-
{{\cal D}_{ijk}}^2(u)]^{-3/2}\,,
\end{eqnarray}
where
${\tilde \Delta}=\beta a\Delta$ and $X_{1...m}$ are determinants of $m\times m$-matrices
\begin{eqnarray}
%\label{Pi}
X_{1...m}=\mbox{det}_{m\times m}\tilde{D}_{l:k}\,,\quad
\tilde{D}_{l:k}=\tilde{{\cal D}}_{i_li_0i_k}\,.
\nonumber
\end{eqnarray}
%%%%%%%%%%%%%%%%%%%%%    BIBLIOGRAPHY  %%%%%%%%%%%%%%%%%%%%%%%%

\newpage
\section*{Figure Captions}

\FigCaption{1}{
Plot of the correlation functions ${\cal R}^{(0,1)}_1$ and the replica
symmetry
breaking parameter $x$ vs the dispersion of disorder $\tilde{\Delta}$ for
polymer with the degree of polymerization $N=50$.
Here ${u}_2\,l^{-3} = -50$, ${u}_3\,l^{-6} = 60$,
${u}_4\,l^{-9} = 48$. The dispersion $\tilde{\Delta}_f$ denotes
the point of the freezing transition.
}
\FigCaption{2}{
Plot of the correlation functions ${\cal R}^{(1)}_q$ (left--hand side
of the picture) and ${\cal R}^{(0)}_q$ (right--hand side)
vs the chain index $q$ for polymer with the degree of polymerization $N=50$.
Here ${u}_2\,l^{-3} = -50$, ${u}_3\,l^{-6} = 60$,
${u}_4\,l^{-9} = 48$.
Lines (a) - (c) correspond respectively to the values of the dispersion of
disorder $\tilde{\Delta} = 2$ (near the freezing transition line), $3$ and $4$.
For convenience we have drawn the plots only on one of the sides,
and the correlation functions may be extended to the other one by the
symmetry property: ${\cal R}^{(0,1)}_q = {\cal R}^{(0,1)}_{N-q}$.
}

\FigCaption{3}{
The phase diagram of a random copolymer in variables of the dispersion,
$\tilde{\Delta}$, and mean value, ${u}_{2}\,l^{-3}$, of the second virial
coeficient. Lines (C) and (F) correspond to collapse and freezing
transitions respectively.
This diagram has been obtained from the data
for a polymer with the degree of polymerization $N = 20$, the third and
the fourth virial coefficients are equal to ${u}_{3}\,l^{-6} = 60$,
${u}_4\,l^{-9} = 48$.
No substantial qualitative changes are observed for larger $N$.
Due to finite width of the collapse transition for finite--size systems
the line (C) is drawn by the maximal descent of the squared radius of
gyration.
}

\FigCaption{4}{
Plots of the phase separation parameter, $\delta R_{g}^{2}\,l^{-2}$, vs the
dispersion of the second virial coeficient, $\tilde{\Delta}$,
for polymer with $N = 20$, ${u}_{3}\,l^{-6} = 60$,
${u}_4\,l^{-9} = 48$, $\lambda_{0} = 0$,
$\Delta = 5$ and the values of second virial coeficient
${u}_{2}\,l^{-3} = -10$ (diamonds), $-40$ (crosses),
$-70$ (quadrangles) and $-100$ (triangles).
}

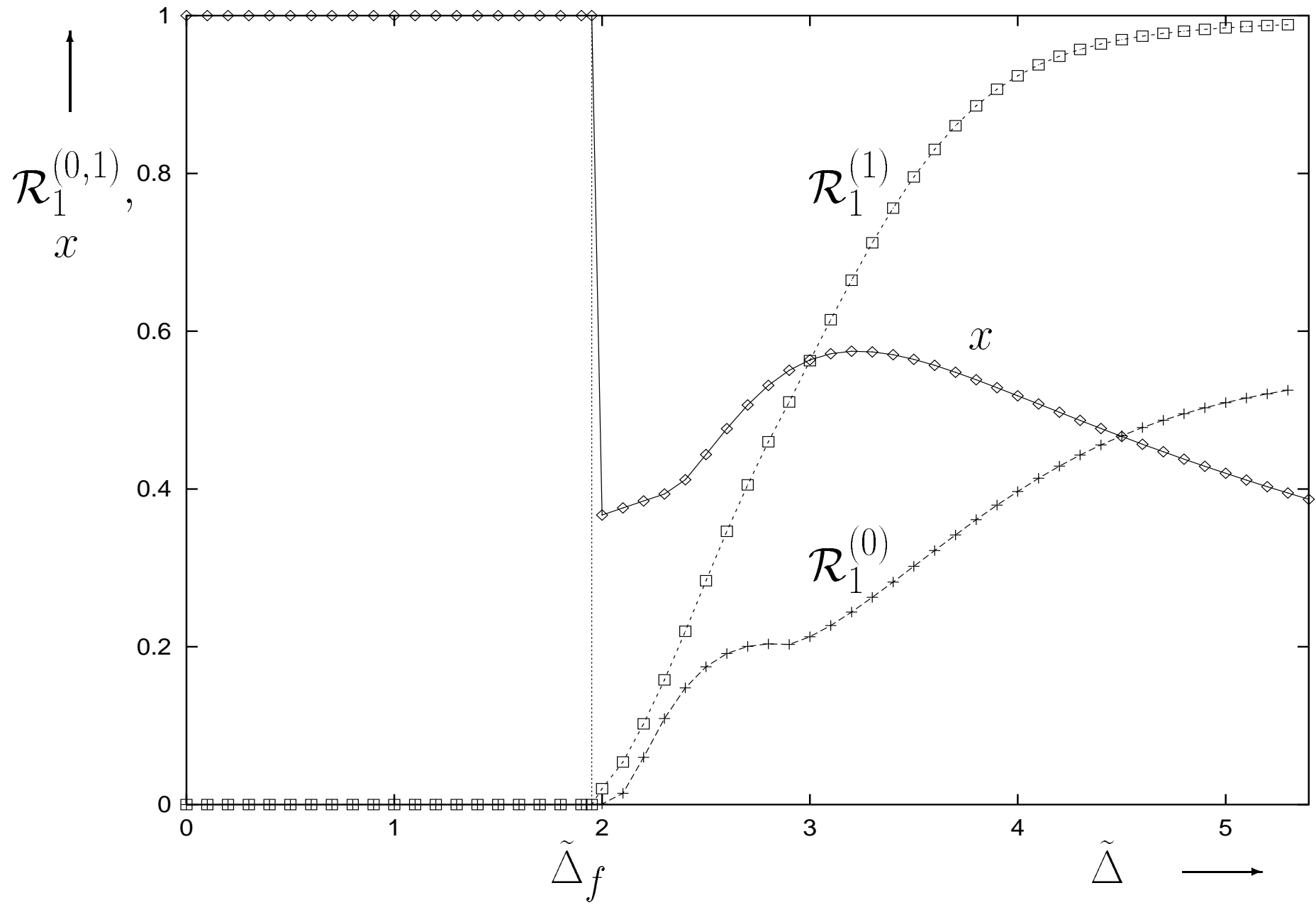
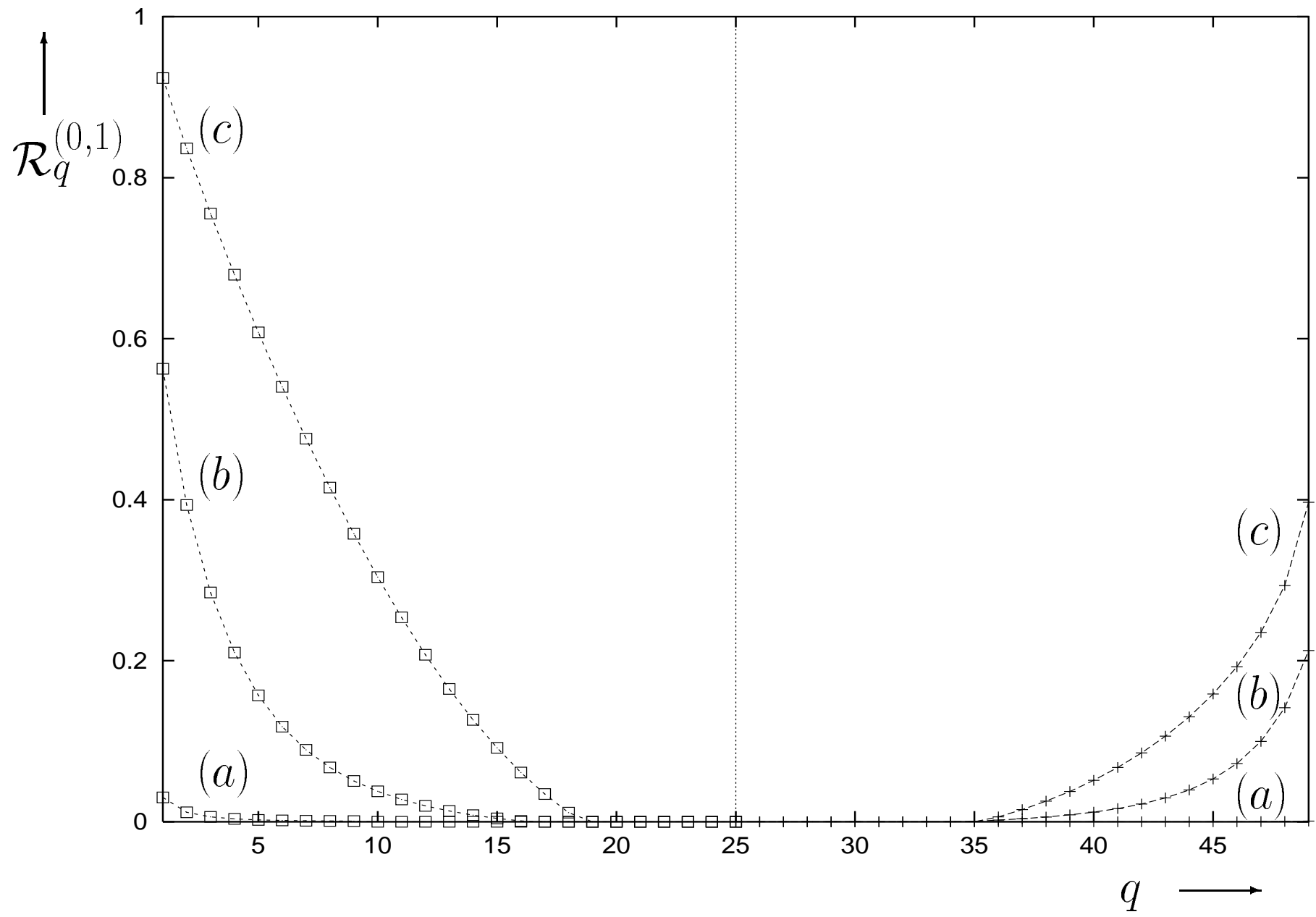
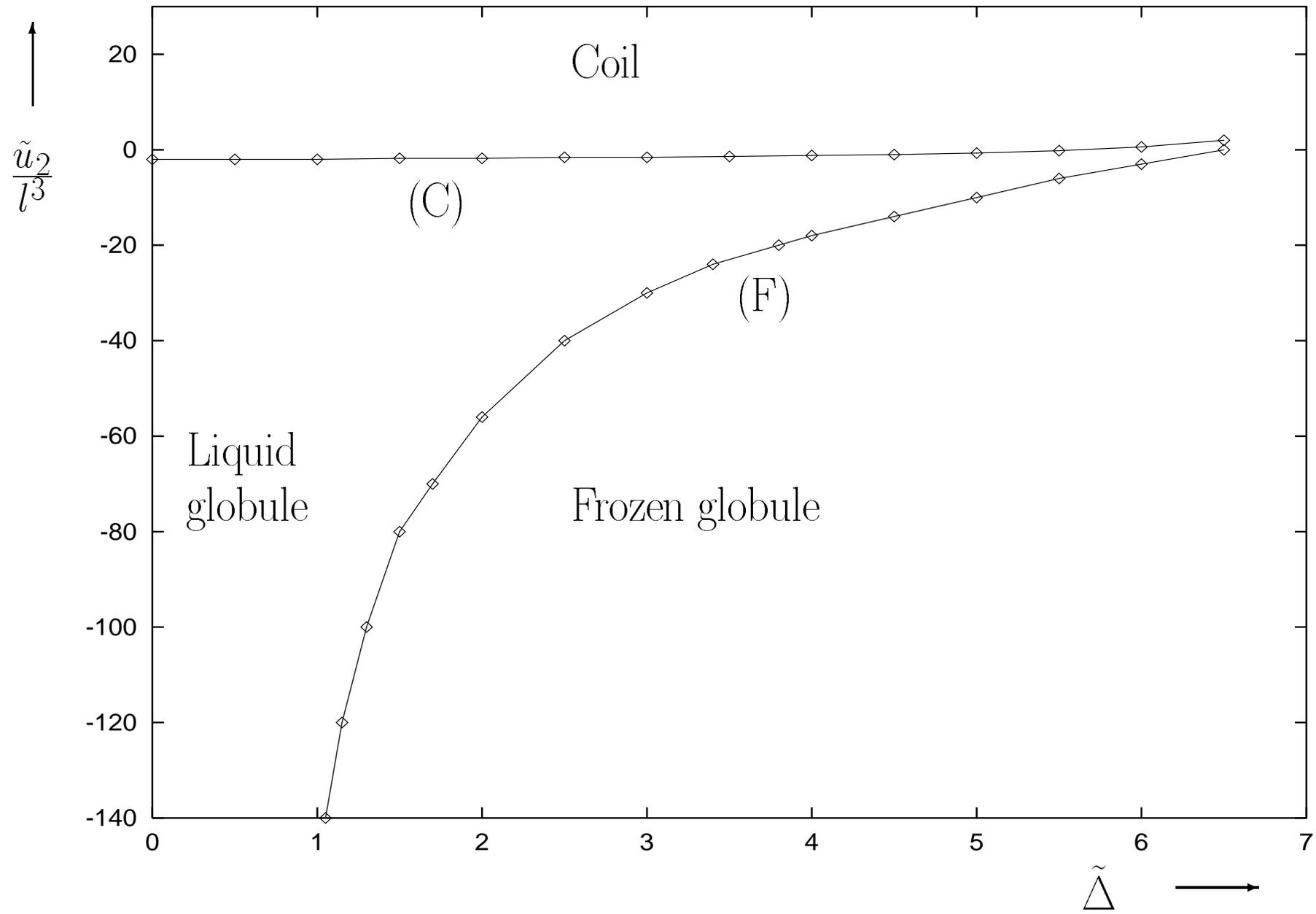
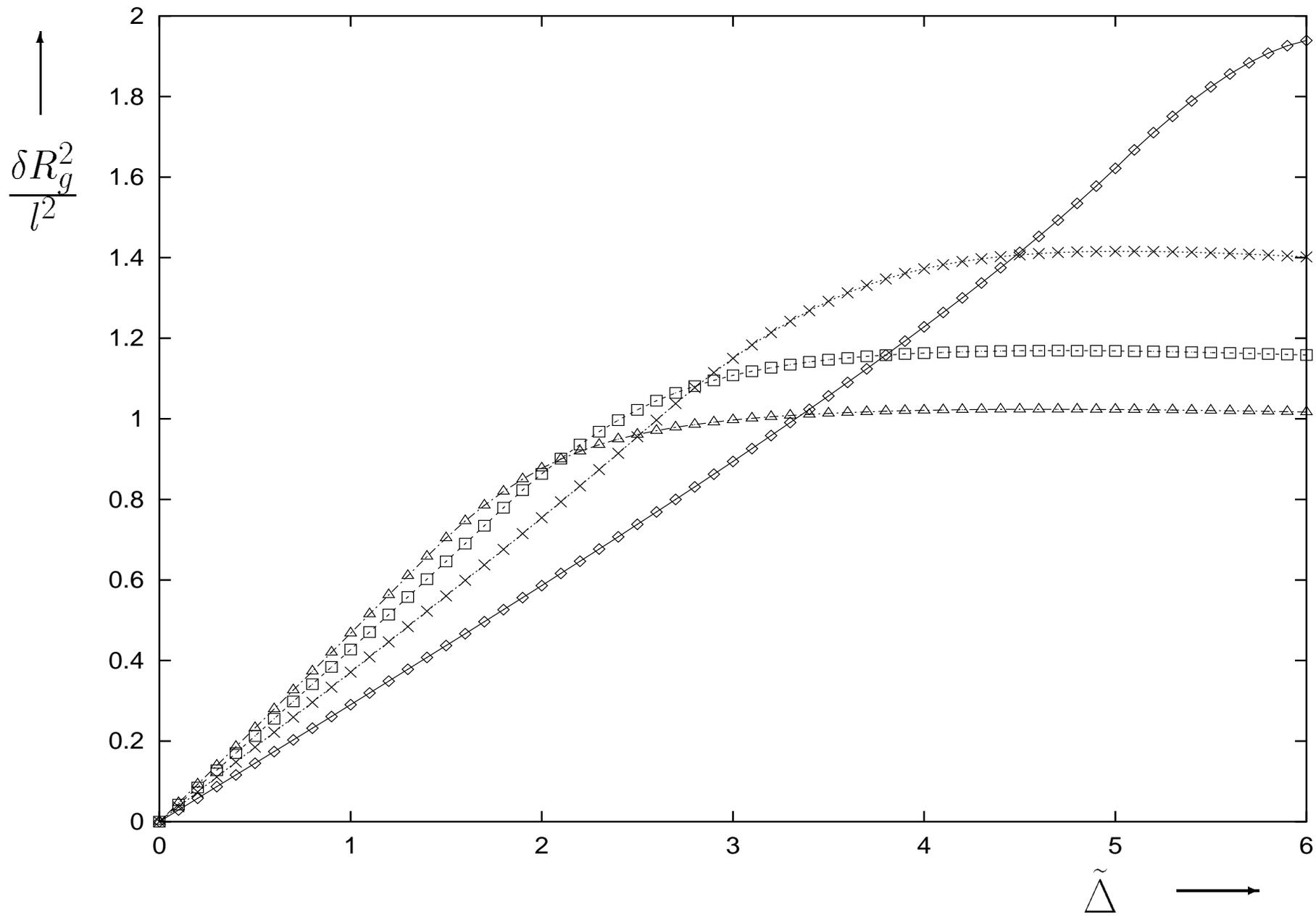
\end{document}